\documentclass{cernrep} 
\usepackage{texnames}
\usepackage[T1]{fontenc}
\begin{document}
\def\GeV{{\rm GeV}}
\title{The role of uncertainties in parton distribution functions}
 
\author{R.S. Thorne$^{1}$\footnote{Royal Society University Research Fellow}}

\institute{Department of Physics and Astronomy, University College London, 
WC1E 6BT, UK}

\maketitle % this produces the title block

\begin{abstract}
I consider the uncertainties in parton distributions and the consequences for 
hadronic cross-sections. There is ever-increasing 
sophistication in the relationship between the uncertainties of the 
distributions and the errors on the experimental data used to extract 
them. However, I demonstrate that this uncertainty is frequently 
subsumed by that due to the choice of data used in fits, and 
more surprisingly 
by the precise details of the theoretical framework used. Variations 
in heavy flavour prescriptions provide striking examples.    
\end{abstract}
 
\section{Introduction}

When calculating cross-sections for scattering 
processes involving hadronic particles one requires detailed knowledge of the 
input parton distributions. The uncertainties in the latter propagate into 
the uncertainties on the former, and are often significant and sometimes 
dominant. 
The parton distributions can be derived within QCD using the 
Factorization Theorem, i.e.  
the cross-section for a physical cross-section at the LHC 
can be written in the factorised form
\begin{equation}
\sigma(pp\to X_P)\propto \sum_i\sum_j C^P_{ij}(x_1,x_2,\alpha_s(M^2))\otimes
f_{i}(x_1,M^2)\otimes f_{j}(x_2,M^2), 
\label{a}
\end{equation}
up to small corrections, where $P$ represents some arbitrary process with
hard scale (e.g. particle mass, jet $E_T$, ...).  
The coefficient functions $C^P_{ij}(x_1,x_2,\alpha_s(M^2))$ 
describing the hard scattering process of the two incoming partons 
are process dependent but calculable as a power-series
in $\alpha_S(M^2)$. The $f_{i}(x,M^2)$ 
are the parton distributions -- heuristically the 
probability of finding a parton of type $i$ carrying a fraction $x$ 
of the momentum of the proton. The parton distributions are 
not calculable from first principles, but evolve with $M^2$ 
in a perturbative manner governed by the splitting functions 
$P_{ij}(x,\alpha_s(M^2))$ which are calculable order by order in 
perturbation theory. Hence, once measured at one scale the
distributions can be predicted 
at other scales. 

In this article I will briefly review the extraction of the parton 
distributions and the resulting uncertainties. This is an update of a previous 
article in this series of Workshops \cite{phystat1}, so 
I will concentrate on new developments. A full discussion of fitting 
procedures and uncertainties due to 
experimental errors on the input data 
is found in \cite{phystat1}, but I will very briefly
restate the essentials, including some updates. 

There are a variety of sets of parton distributions which are obtained 
by a comparison to all available data (so-called global fits) 
\cite{MRST04,CTEQ65} or to smaller subsets of mainly structure function 
data  \cite{Alekhin, ZEUSJ,H1}, sometimes only in the nonsinglet sector
\cite{blumlein,neural}. All follow the same general principle.  
The fit usually proceeds by starting the parton evolution at a low 
scale $Q_0^2$ and evolving partons upwards 
(sometimes also downwards) using fixed order evolution equations. 
The default has long been next-to-leading order (NLO), but the 
next-to-next-to-leading order
(NNLO) splitting functions were recently calculated \cite{NNLOs},
and sets of NNLO distributions are also available \cite{AlekhinNNLO, MRST06}.
In principle, there are 11 different parton 
distributions (assuming isospin symmetry and ignoring the top quark) -- 
the 5 quarks, up, down, strange, charm, and bottom and their antiquarks, 
and the gluon distribution. Until recently these were not all considered 
independent, but there is now some evidence for asymmetry between strange 
quarks and 
antiquarks \cite{NuTeVdimuon}, and moreover all quarks evolve slightly 
differently from their antiquarks due to evolution effects which begin 
at NNLO. However, in  
practice $m_c, m_b \gg \Lambda_{{\rm QCD}}$, so the heavy 
parton distributions are usually determined perturbatively and there are  
7 independent input parton sets, each parameterised in a particular form, e.g.
\begin{equation}
xf(x, Q_0^2) = A(1-x)^{\eta}(1+\epsilon x^{0.5}+\gamma x)
x^{\delta}.
\end{equation}  
The partons are constrained by a number of sum rules:
i.e. conservation of the number of valence up and down quarks, 
zero number asymmetry for the other quarks and the 
conservation of the momentum carried by partons.
The last is an important constraint on the form of the gluon, which is only 
probed indirectly. 
In determining partons one needs to consider that not only are there 
many different distributions, but there is also a wide distribution of
$x$ from $0.75$ to $0.00003$. 
One needs many different types of experiment for
full determination, as discussed in \cite{phystat1}. For instance, the MRST 
(now MSTW \cite{MSTW}) group use 29 different types of data set.   

The quality of the fit is determined by the $\chi^2$ of the fit to 
data, which may be calculated in various ways. 
The simplest is to add statistical and systematic errors in quadrature, 
which ignores correlations between data points, but is sometimes 
quite effective. Also, the information on the data often 
means that only this method is available.
More properly one uses the full covariance matrix which 
is constructed as
\begin{equation}
C_{ij} = \delta_{ij} \sigma_{i,stat}^2 + \sum_{k=1}^n \rho^k_{ij}
\sigma_{k,i}\sigma_{k,j}, \qquad 
\chi^2 = \sum_{i=1}^N\sum_{j=1}^N (D_i-T_i(a))C^{-1}_{ij}
(D_j-T_j(a)),
\label{k}
\end{equation}
where $k$ runs over each source of correlated systematic error,
$\rho^k_{ij}$ are the correlation coefficients,
$N$ is the number of data points, $D_i$ is the 
measurement and $T_i(a)$ is the theoretical prediction depending on 
parton input parameters $a$. An alternative that produces identical results 
if the errors are small is to 
incorporate the correlated errors into the theory prediction
\begin{equation}
f_i(a,s) = T_i(a) + \sum_{k=1}^n s_k \Delta_{ik}, \qquad \chi^2 = 
\sum_{i=1}^N \biggl(\frac{D_i-f_i(a,s)}{\sigma_{i,unc}}
\biggr)^2 + \sum_{k=1}^n s_k^2,
\label{m}
\end{equation}
where $\Delta_{ik}$ is the one-sigma correlated error for point 
$i$.
One can solve analytically for the $s_k$ \cite{lmethod}. 

\begin{figure}
\vspace{-0.6cm}
\centering\hspace{-0.8cm}\includegraphics[width=.45\linewidth]
{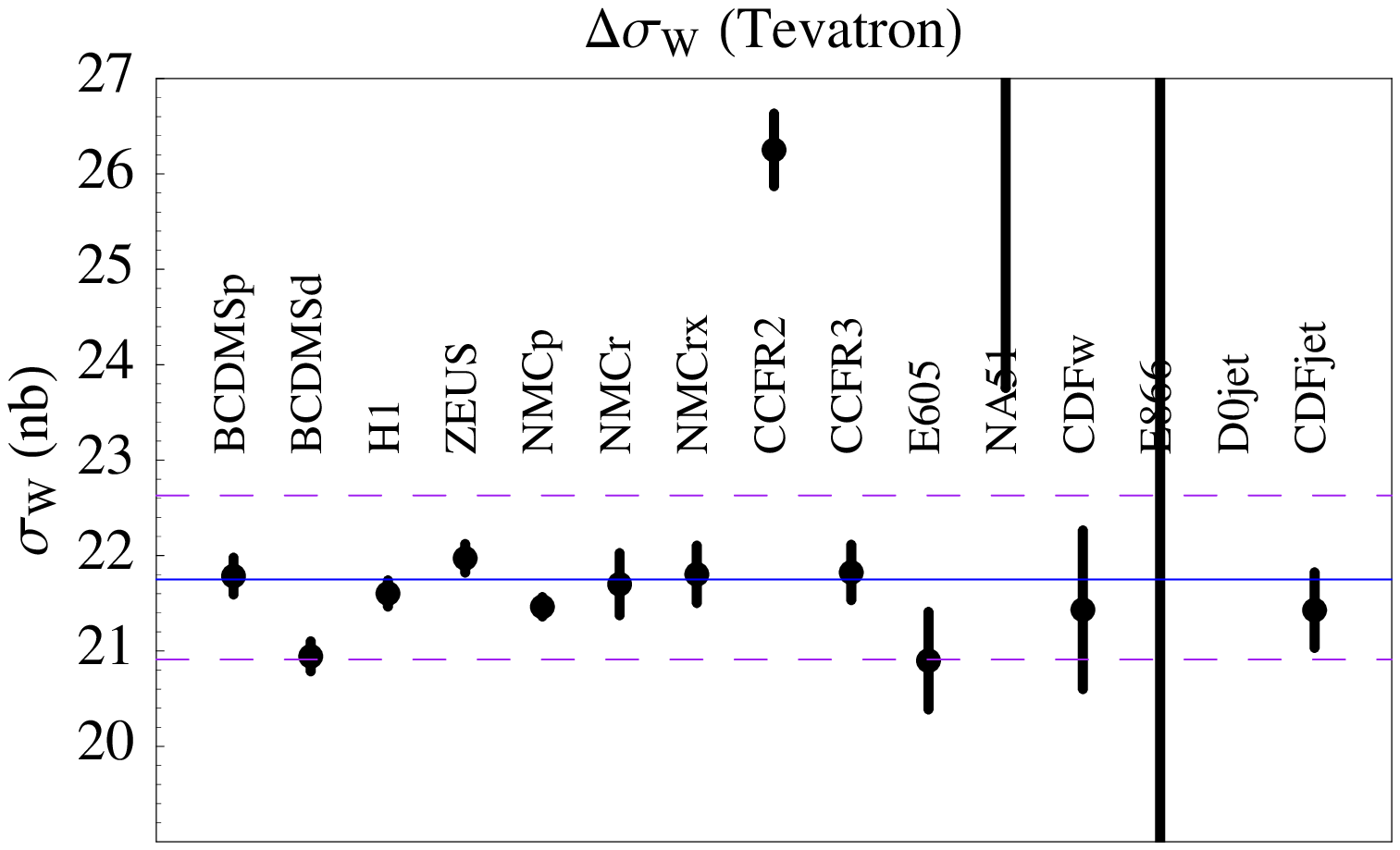}
\includegraphics[width=.45\linewidth]{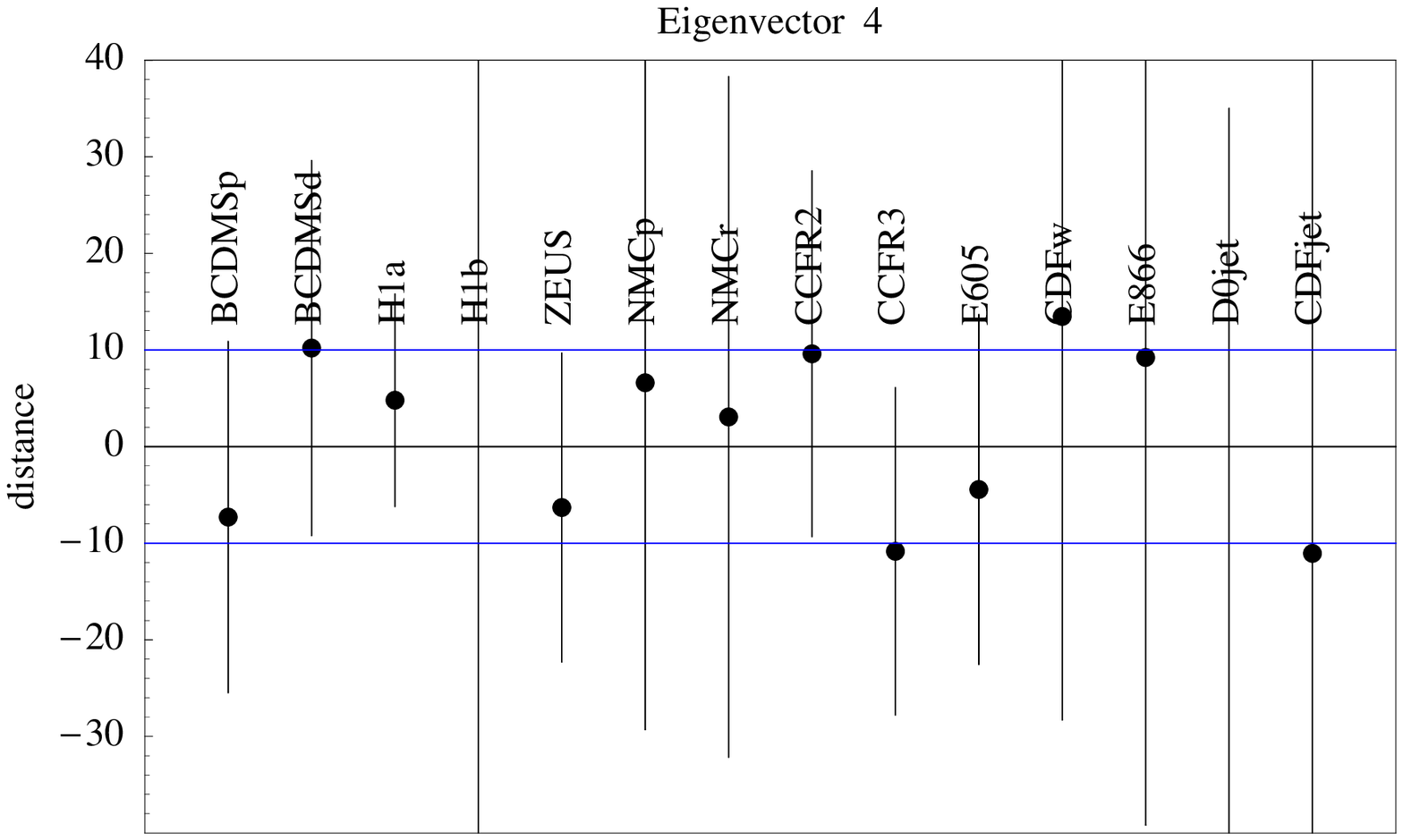}   
\vspace{-0.2cm}
\caption{The best value of $\sigma_W$ and the uncertainty using 
$\Delta \chi^2=1$
for each data set in the CTEQ fit (left) and  the $90\%$ confidence limits
for each data set as a function of $\sqrt{\Delta \chi^2=100}$ 
for one particular eigenvector}
  \label{cteq1sigma}
\vspace{-0.3cm}
\end{figure}

Having defined the fit quality
there are a number of different approaches for obtaining parton uncertainties. 
The most common is the Hessian (Error Matrix) approach. 
One defines the Hessian matrix by
\begin{equation}
\chi^2 -\chi_{min}^2 \equiv \Delta \chi^2 = \sum_{i,j} 
H_{ij}(a_i -a_i^{(0)})
(a_j -a_j^{(0)}).
\label{r}
\end{equation}
One can then use the standard formula for linear error propagation:  
\begin{equation}
(\Delta F)^2 = \Delta \chi^2 \sum_{i,j} \frac{\partial F}
{\partial a_i}(H)^{-1}_{ij}  
\frac{\partial F}{\partial a_j}.
\label{t}
\end{equation}
This has been used to find partons with errors by $H1$ \cite{H1}
and Alekhin \cite{Alekhin}. In practice it is problematic due to extreme 
variations in $\Delta \chi^2$ in different directions in parameter space.
This is improved by finding and rescaling the eigenvectors of $H$, 
a method developed by CTEQ \cite{cteqpap1,hmethod}, and now used by
most groups. The uncertainty on a physical quantity is 
\begin{equation}
(\Delta F)^2 = \sum_{i} \bigl(F(S_i^{(+)})-F(S_i^{(-)})\bigr)^2,
\label{v}
\end{equation}
where $S_i^{(+)}$ and $S_i^{(-)}$ are PDF sets 
displaced along eigenvector
directions by the given $\Delta \chi^2$. 

One can also investigate 
the uncertainty on a given physical quantity using the 
Lagrange Multiplier method, first suggested by CTEQ \cite{lmethod} 
and also used by MRST \cite{MRSTerror1}. One performs 
the global fit while constraining the value of some physical 
quantity, i.e.  minimise 
\begin{equation}
\Psi(\lambda,a) = \chi^2_{global}(a)  + \lambda F(a)
\label{ac}
\end{equation}
for various values of $\lambda$. This gives the set of best fits for 
particular values of the parameter $F(a)$ without relying on the quadratic 
approximation for $\Delta\chi^2$, but has to be done anew for each quantity.  

In each approach there is uncertainty in choosing 
the ``correct'' $\Delta \chi^2$. In principle this should be one unit, 
but given the complications of a full global fit this gives unrealistically 
small uncertainties. This can be seen in the left of Fig.~\ref{cteq1sigma}
where the variation in the predictions for $\sigma_W$ using $\Delta \chi^2 =1$
for each data set has an extremely wide scatter compared to the uncertainty. 
CTEQ choose $\Delta \chi^2 \sim 100$ \cite{lmethod}. The $90\%$ confidence 
limits for the fits to the larger individual data sets when 
$\sqrt{\Delta \chi^2}$ in the CTEQ fit is increased by a 
given amount are shown in 
the right of Fig.~\ref{cteq1sigma}. As one sees, a couple of sets may be some 
way beyond their $90\%$ confidence limit for $\Delta \chi^2 = 100$. The 
MRST/MSTW  group chooses $\Delta \chi^2 = 50$
to represent the $90\%$ confidence limit for the fit. Other groups with much 
smaller data sets and fewer complications still use $\Delta \chi^2=1$.

\begin{figure}
\vspace{-0.6cm}
\centering\hspace{-0.8cm}\includegraphics[width=.45\linewidth]
{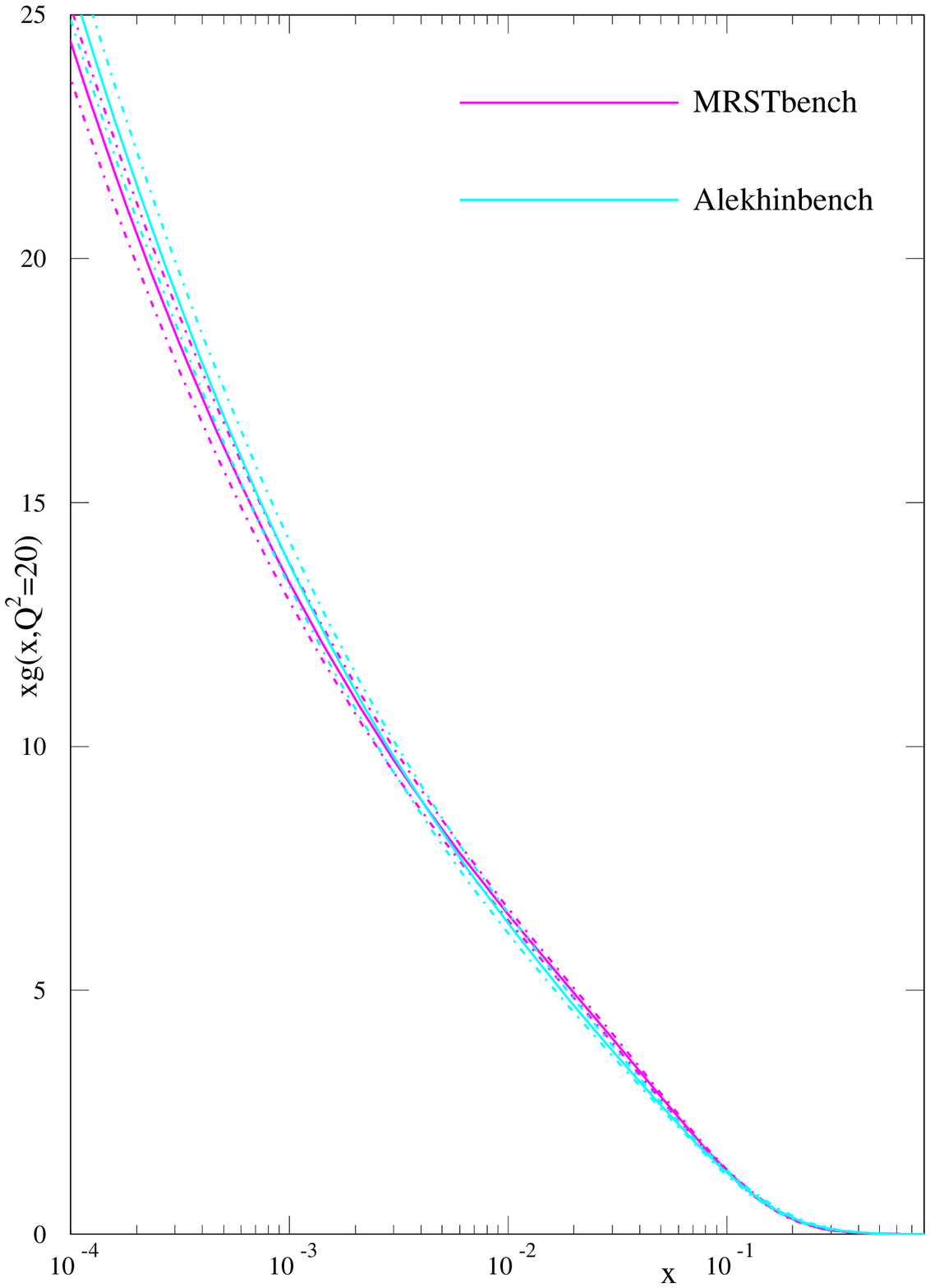}
\includegraphics[width=.45\linewidth]{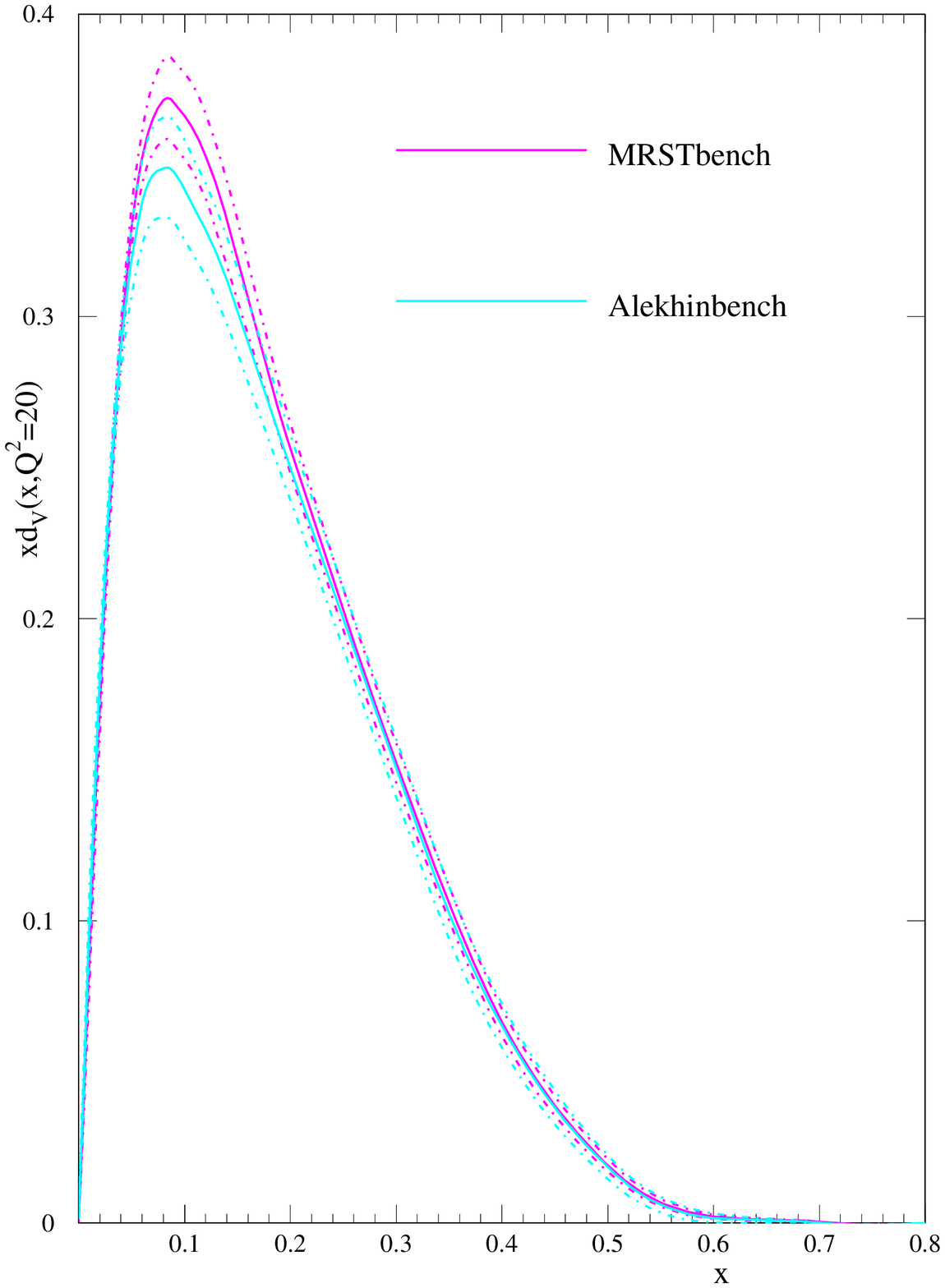}   
\vspace{-0.2cm}
\caption{Comparison of the benchmark gluon distributions and  
$d_V$ distributions}
  \label{glucompsa}
\vspace{-0.3cm}
\end{figure}

There are other approaches to finding the uncertainties. In the offset method
the best fit is obtained by minimising the $\chi^2$ using
only uncorrelated errors. The systematic errors on the parton parameters 
$a_i$ are determined by letting each $s_k = \pm 1$ and adding 
the deviations in quadrature. 
This method was used in early H1 fits \cite{pascaud} and by 
early ZEUS fits \cite{mbfit}, but is uncommon now. There is also the 
statistical approach used by Neural Network group \cite{neural}. Here one 
constructs a set of Monte Carlo replicas $\sigma^k(p_i)$ of the original 
data set  $\sigma^{data}(p_i)$ which gives a
representation of $P[\sigma(p_i)]$ at points $p_i$. 
Then one trains a neural network for the parton distribution function on each 
replica, obtaining a representation of the pdfs $q_i^{(net)(k)}$.
The set of neural nets is a representation of the probability 
density -- i.e. the mean $\mu_O$ and deviation 
$\sigma_O$ of an observable $O$ is given by
\begin{equation}
\mu_O = \frac {1}{N_{rep}}\sum_1^{N_{rep}}O[q_i^{(net)(k)}], \quad
\sigma_O^2 =\frac {1}{N_{rep}} \sum_{1}^{N_{rep}}(O[q_i^{(net)(k)}]-\mu_O)^2.
\end{equation}
One can incorporate full information about measurements and their error 
correlations in the distribution of $\sigma^{data}(p_i)$. 
This is does not rely on 
the approximation of linear propagation of errors 
but is more complicated and time intensive. It is currently done for the 
nonsinglet sector only.

\section{Sources of Uncertainty}

In recent years there has been a great deal of work on the correct and 
complete inclusion of the experimental errors on the data when extracting 
the partons and their uncertainties. However,
to obtain a complete estimate of errors, one also 
needs to consider the effect of the decisions and assumptions made when 
performing the fit, e.g. cuts made on the data, data sets fit 
and parameterization for the input sets. 

\begin{figure}
\vspace{-0.0cm}
\centering\hspace{-0.8cm}\includegraphics[width=.45\linewidth]
{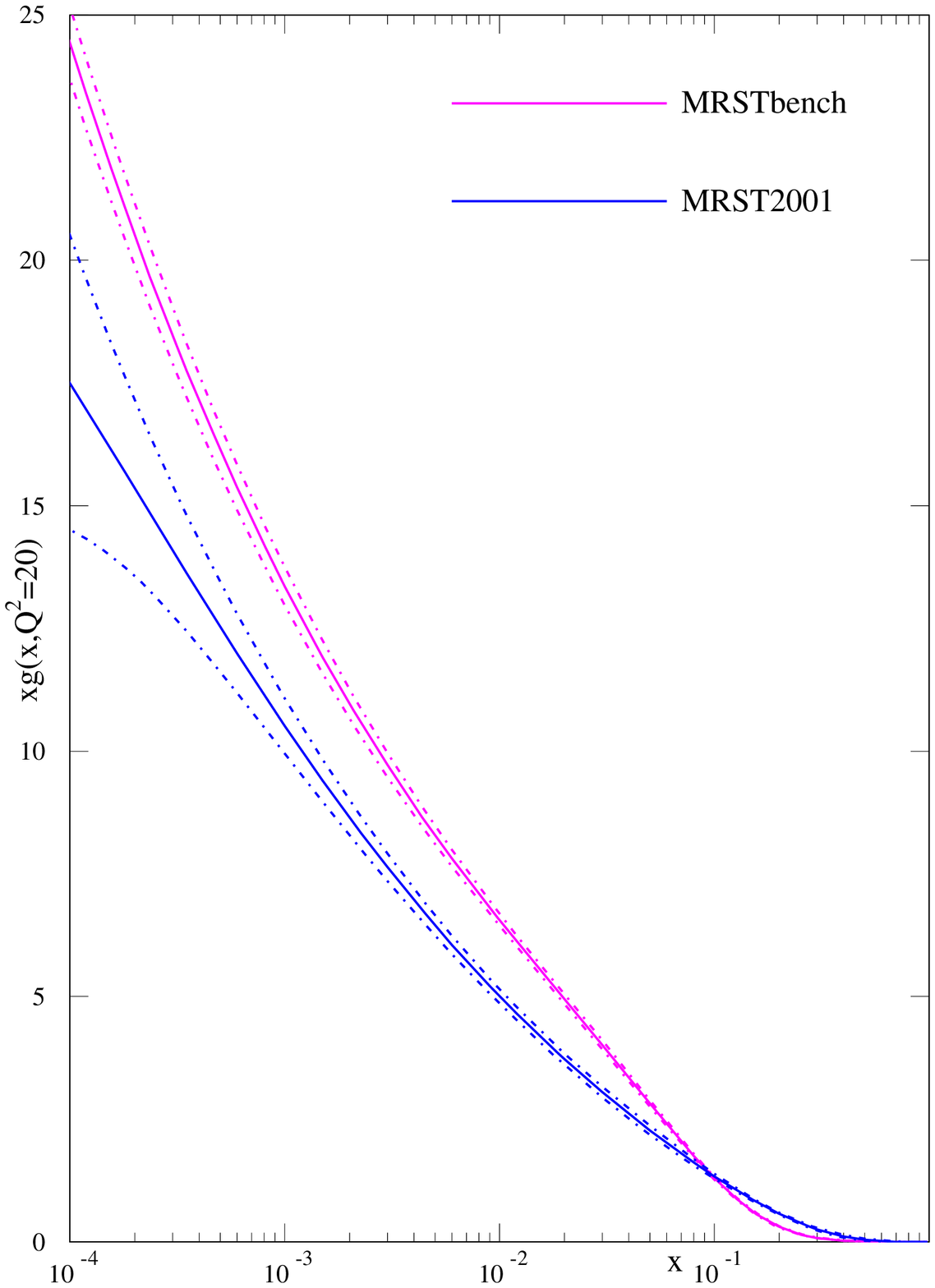}
\includegraphics[width=.45\linewidth]{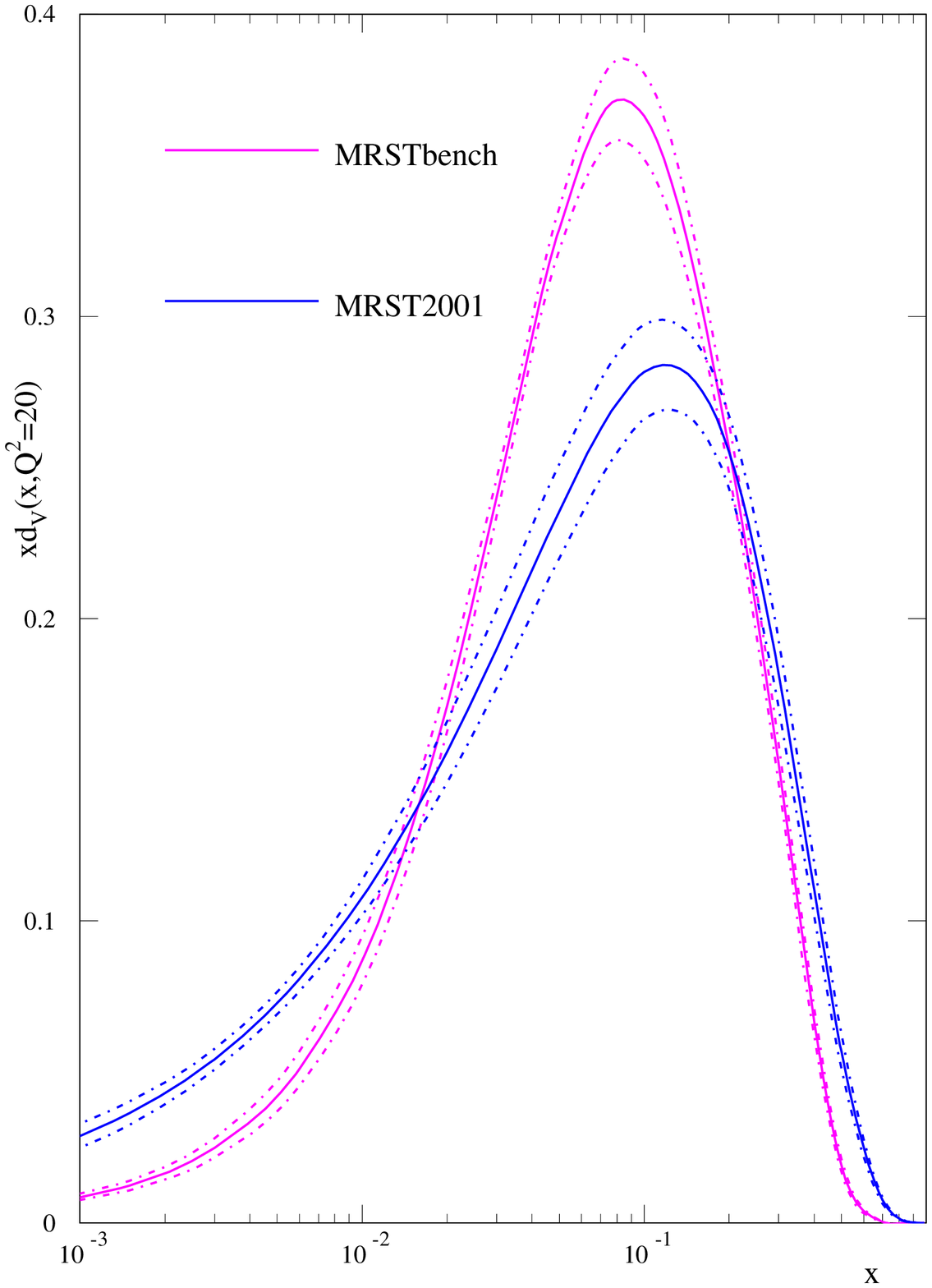}   
\vspace{-0.2cm}
\caption{Comparison of the benchmark gluon and  
$d_V$ distribution with the corresponding MRST2001E partons} 
  \label{glucomps}
\vspace{-0.4cm}
\end{figure} 

As an exercise for the HERA-LHC \cite{HERALHC} 
workshop, partons were produced from fits to some sets of structure function 
data for $Q^2 > 9 \GeV^2$ using a common form of parton inputs at 
$Q_0^2 =1 \GeV^2$. Partons were 
obtained using the rigorous treatment of all systematic errors 
(labelled Alekhin)
and using the simple quadratures approach (labelled MRST), both using 
$\Delta \chi^2=1$ to define the limits of uncertainty. 
This benchmark test is clearly a very conservative approach 
to fitting that should give reasonable partons with bigger than normal 
uncertainties. As seen in  Fig.~\ref{glucompsa} there are 
small differences in the central values and similar errors, i.e.
the two sets are fairly consistent. It is more interesting to 
compare the HERA-LHC benchmark partons to partons obtained from a global fit
\cite{MRSTerror1}, where the uncertainty is 
determined using $\Delta \chi^2=50$. There is an 
enormous difference in the central values,
sometimes many $\sigma$, as seen in Fig.~\ref{glucomps}, 
although the uncertainties are similar using 
$\Delta \chi^2=1$ compared to 
$\Delta \chi^2=50$ with approximately twice the data. 
Moreover, $\alpha_S(M_Z^2)\!=\!0.1110\pm 0.0015$
from the benchmark fit compared to $\alpha_S(M_Z^2)\!=\!0.119\pm 0.002$. 
Something is clearly seriously wrong in one of these analyses, and 
indeed partons from the benchmark fit fail when compared 
to most data sets not included. This implies that partons should be 
constrained by all possible reliable data. 

%\begin{figure}
%\vspace{-0.7cm}
%\centering\hspace{-0.8cm}\includegraphics[width=.45\linewidth]{badglu.ps}   
%\vspace{-1cm}
%\caption{Comparison of various gluon distributions} 
%  \label{badglu}
%\vspace{-0.4cm}
%\end{figure} 

\begin{figure}
\vspace{-0.7cm}
\centering\hspace{-0.8cm}\includegraphics
[width=.45\linewidth]{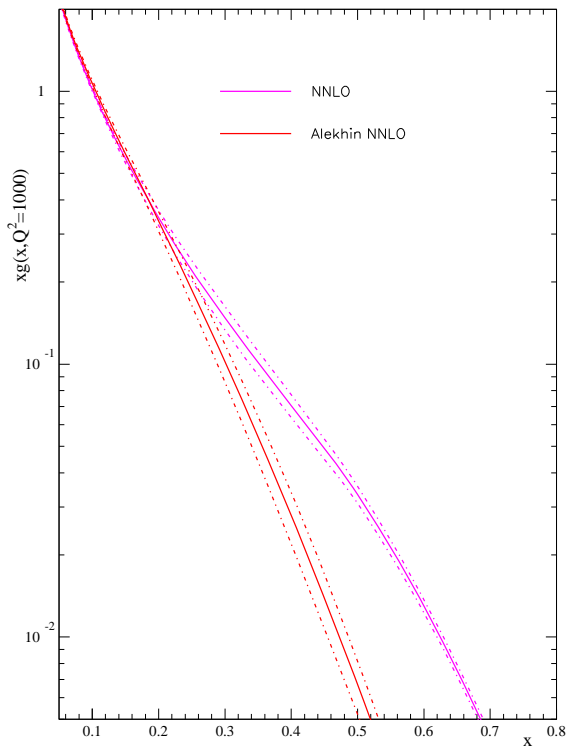}   
\vspace{-0.2cm}
\caption{Comparison of MRST and Alekhin NNLO gluon distributions at high $x$} 
  \label{glualcomph}
\vspace{-0.4cm}
\end{figure} 
 
The benchmark partons above are not a realistic set of partons, but similar 
examples are found when comparing different sets of published parton 
distributions. For example, the valence quarks extracted from the 
nonsinglet analysis in \cite{blumlein} (see Figs. 9 and 10) are 
different from a variety of alternatives by much more than the uncertainties.
Indeed, various gluon distributions, all obtained by fitting to 
small $x$ HERA data \cite{H1,HERA} are  
very  different despite what is meant to be the main constraint on the data 
being the same in each case. It is particularly illustrative to look at the 
difference in the high-$x$ gluons of MRST and Alekhin in Fig.~\ref{glualcomph}.
This is for NNLO, but is similar at NLO.
Here the difference above $x=0.2$ is a large factor, and very much bigger 
than each
uncertainty (calculated using $\Delta \chi^2 =1$ for Alekhin and  
$\Delta \chi^2 =50$ for MRST.) It seems that the HERA data require  
a gluon distribution for the 
very best fit which is incompatible with the Tevatron jet 
data \cite{jets}, and the standard error analysis does not accommodate  
this. As a further point, at NNLO one of the few hard cross-sections required 
in a global fit which is not fully known is that for the jet cross-section. 
It might be argued that one should leave the data out rather than rely on 
the NLO
hard cross-section, as done by MRST. However, this correction is very likely
to be $\sim 5\%$, whereas the change in the gluon distribution if the 
data are left out can be $> 100\%$. This implies, 
to the author at least, that it
is better to include a data set relying on a slight approximation than to 
leave it out and obtain partons which are completely incompatible with it. 

\begin{figure}
\vspace{-0.7cm}
\centering\hspace{-0.8cm}\includegraphics[width=.42\linewidth]
{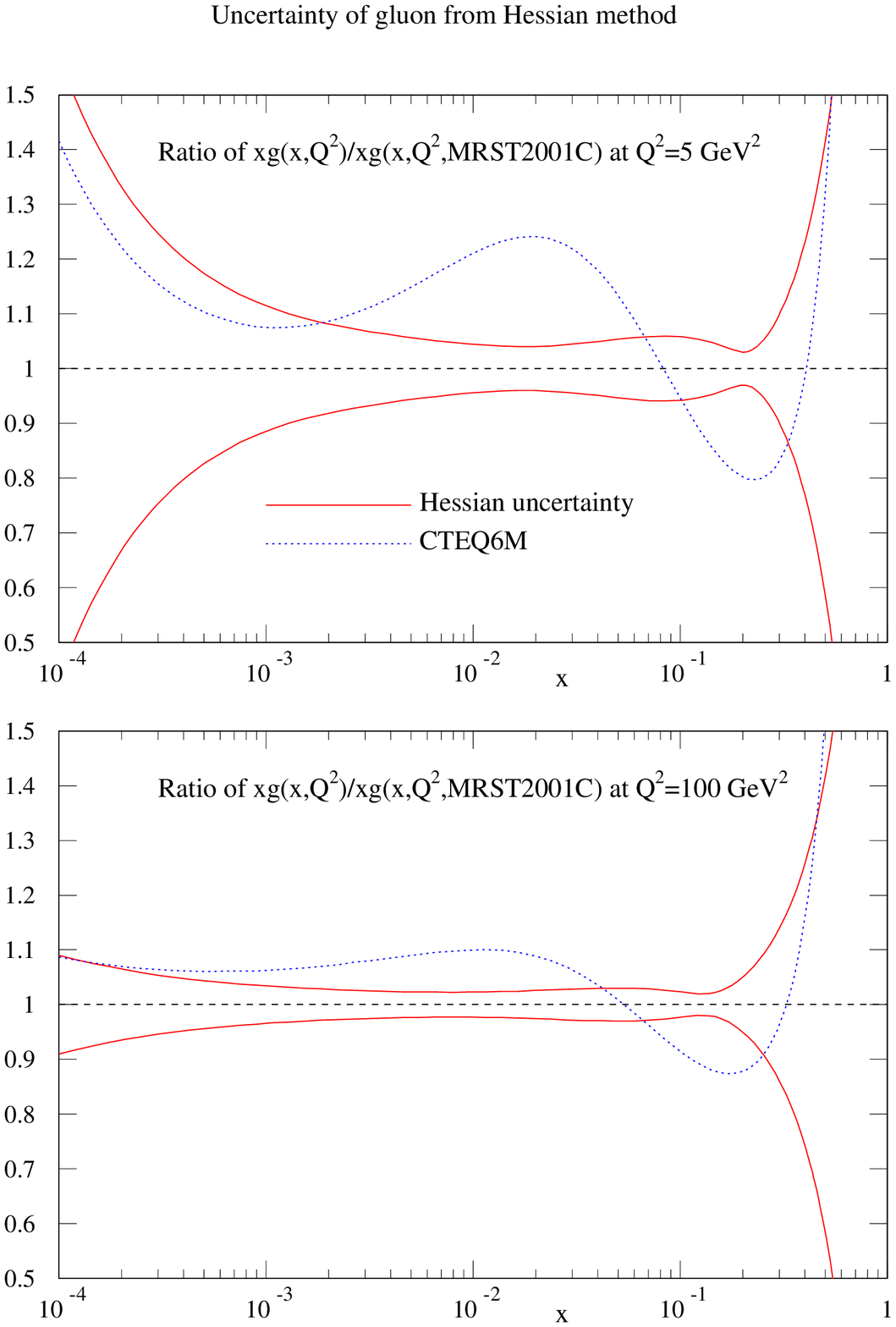}
\hspace{0.5cm}
\includegraphics[width=.42\linewidth]{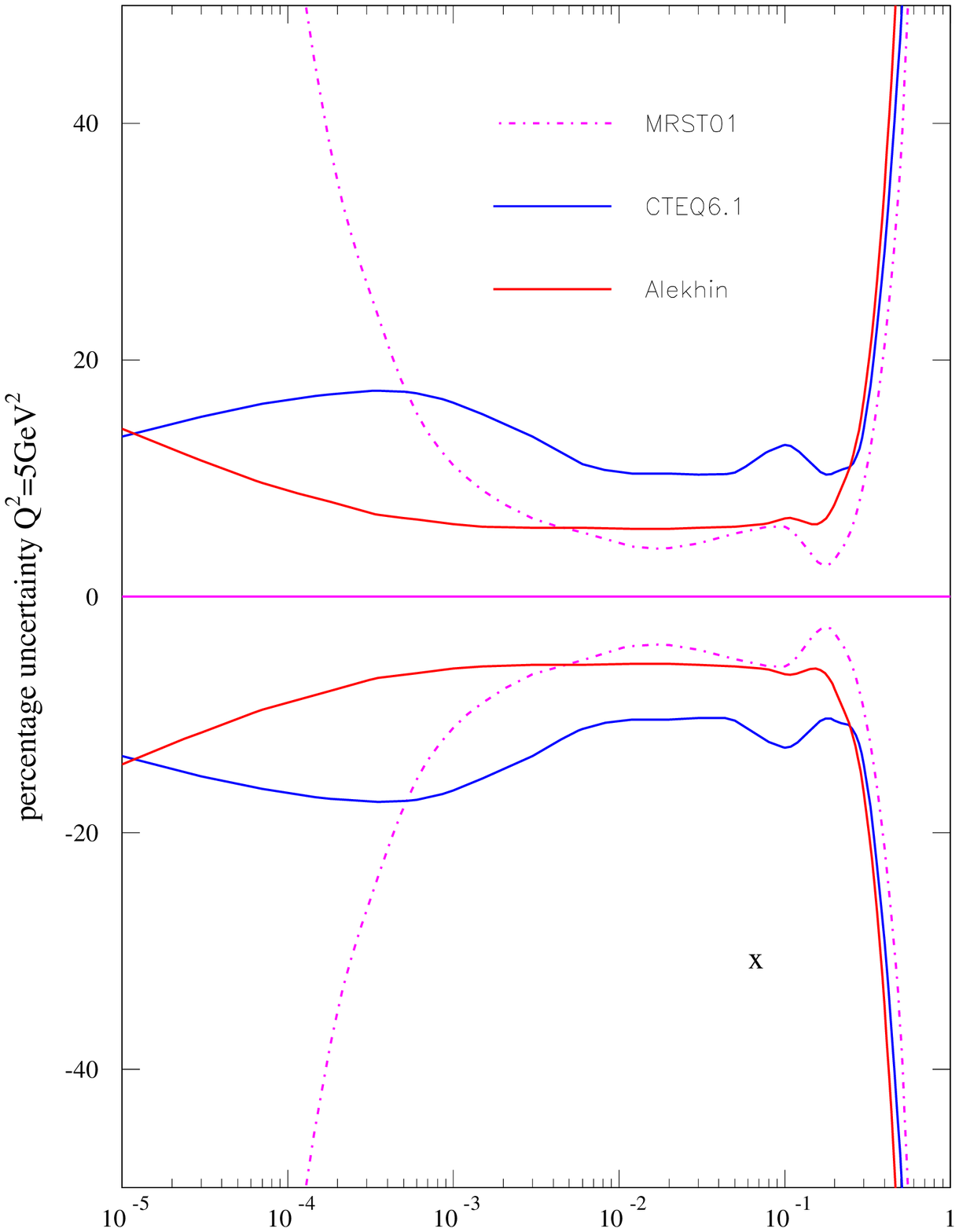}   
\vspace{-0.3cm}
\caption{The MRST gluon distribution with percentage 
uncertainties, and the central CTEQ distribution (left)
and the uncertainties on the MRST, CTEQ and Alekhin
gluon distributions at $Q^2=5\GeV^2$(right)} 
  \label{hessratg}
\vspace{-0.1cm}
\end{figure} 

Even when similar data sets are fit, there can still be significant 
differences in parton distributions and their predictions.The prediction for 
$\sigma_W$ at NLO at the LHC using CTEQ6.5 partons is $202\pm 9$ nb and using
MRST04 partons is $190\pm 5$ nb. This is despite the rather similar data sets 
and  procedures used in the two fits. The different predictions are easily
explained by looking at the left of 
Fig.~\ref{hessratg}. The CTEQ gluon is much bigger 
than MRST at small $x$ and drives quark evolution to be larger. This 
difference is not fully understood but is probably partially due to the fact 
the MRST have lower $Q^2$ cuts on the structure function data, and also 
due to the 
different input parameterisations for the gluon. MRST allow their gluon to 
be negative at small $x$ at input ($Q_0^2=1\GeV^2$) while the CTEQ gluon 
is positive at small $x$ input ($Q_0^2=1.69\GeV^2$), but is very small indeed.
(Further analysis suggests a slightly negative input gluon is preferred, but
only barely \cite{cteqstab} so the freedom is not introduced.)  

%\begin{figure}
%\vspace{-0.7cm}
%\centering\hspace{-0.8cm}\includegraphics
%[width=.43\linewidth]{lowxgluerr.ps}   
%\vspace{-0.2cm}
%\caption{The uncertainties on the MRST, CTEQ and Alekhin
% gluon distributions at $Q^2=5\GeV^2$} 
%  \label{lowxgluerr}
%\vspace{-0.4cm}
%\end{figure} 

The parameterization can have an even more dramatic effect on the uncertainty
than on the central value. The uncertainties on the gluon 
distributions for MRST, CTEQ and Alekhin are shown in the right of 
Fig.~\ref{hessratg}.
One would expect the uncertainty to increase significantly at very small
$x$ as constraints die away. This happens for the MRST gluon. 
The Alekhin gluon does not have as much freedom, but is input at higher scales 
and behaves like $x^{-\lambda}$ at small $x$. The uncertainty is 
due to the uncertainty in $\lambda$ (the situation is similar for ZEUS and H1 
partons).  The CTEQ input gluon behaves like $x^{\lambda}$ at 
small $x$ where $\lambda$ is large and positive. 
The small-$x$ input gluon is tiny and has a very small absolute error. At 
higher $Q^2$ all the uncertainty is due to evolution driven by the higher-$x$, 
well-determined gluon. The very small $x$ gluon no more uncertain than at
$x=0.01-0.001$.

\begin{figure}
\vspace{-0.7cm}
\centering\hspace{-0.8cm}\includegraphics
[width=.45\linewidth]{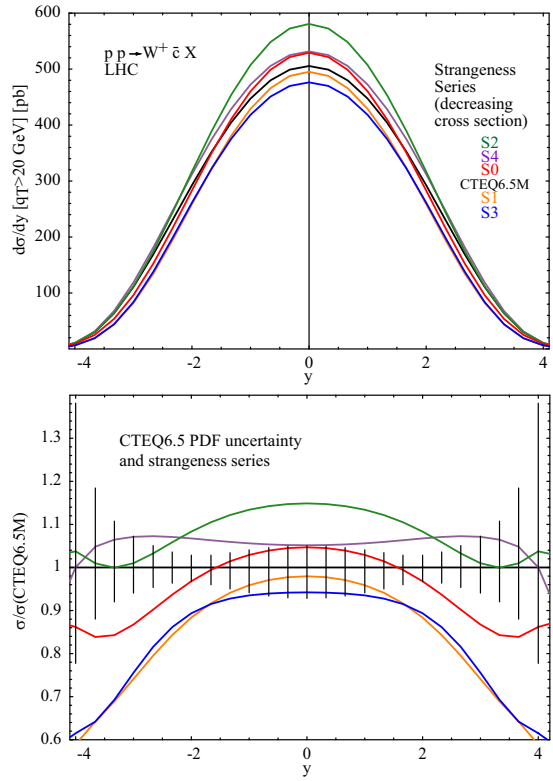}   
\vspace{-0.2cm}
\caption{Uncertainty of predictions for  $W^++\bar c$} 
  \label{CTEQWPC}
\vspace{-0.4cm}
\end{figure} 

Another important source of uncertainty only now becoming clear is due to 
the strange distribution. Until recently this was taken to be a 
fixed and constant fraction of the total sea quark distribution. This did not
allow any intrinsic uncertainty on the strange quark. It is now being fit 
more directly by comparison to dimuon data in neutrino scattering 
\cite{NuTeVdimuon}. In the MSTW fits \cite{MSTW} this results in an increased
uncertainty on all sea quarks since allowing the strange to vary 
independently gives the up and down quarks 
more freedom. CTEQ have 
produced specific parton sets with fits to the strange quark 
\cite{CTEQstrange},  and in Fig.~\ref{CTEQWPC} we see predictions from these 
for production of $W^++\bar c$. CTEQS0 represents the best fit when 
the strange is fit directly. Worryingly, this  can be outside 
the uncertainty band for the default set.  

\section{Theoretical Uncertainties} 
 
\begin{figure}
\vspace{-0.7cm}
\centering\hspace{-0.8cm}\includegraphics
[width=.45\linewidth]{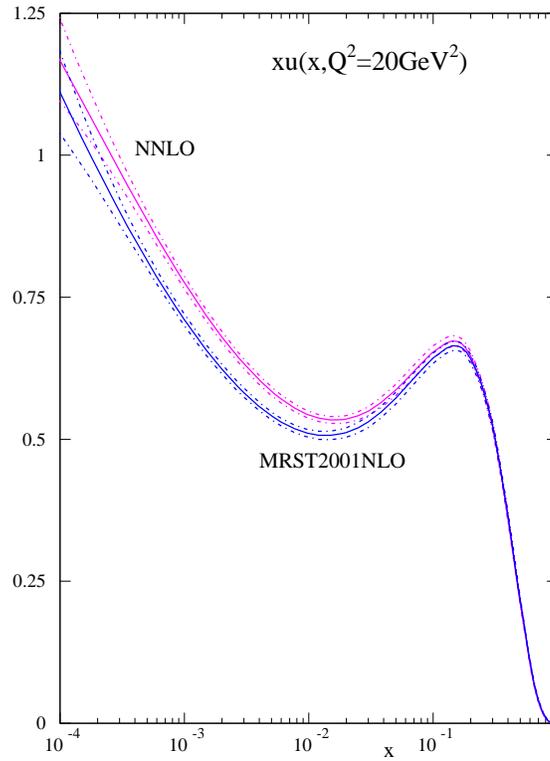}   
\vspace{-0.2cm}
\caption{Comparison between the NLO and NNLO up quark distribution} 
  \label{unnlocomps}
\vspace{-0.4cm}
\end{figure} 

Even if we had an unambiguous definition for the parameterization and the data
sets and cuts used, there would still be additional uncertainties due to the
limited accuracy of the theoretical calculations. 
The sources of theoretical error include higher twist at low scales and
higher orders in $\alpha_S$, and it now seems likely that there may be sizable 
corrections from higher order electroweak corrections at the LHC (see e.g. 
\cite{electroweak}), due to 
$\alpha_W \ln^2(E^2/M_W^2)$ terms in the expansion. 
The higher order QCD errors are due not only to  
fixed order corrections, but also to enhancements at 
large and small $x$ because of terms of the form 
$\alpha_s^n \ln^{n-1}(1/x)$ and $\alpha_s^n 
\ln^{2n-1}(1-x)$ in the perturbative expansion.  
This means that renormalization and factorization 
scale variation are not a reliable way of estimating higher 
order effects because a scale variation at one order will not give any 
indication of an extra
$\ln(1/x)$ or $\ln(1-x)$ at higher orders. Hence,
in order to investigate the true theoretical error we must consider 
some way of performing correct large and small $x$ resummations,
and/or use what we already know about going to higher orders. 

We are now able to look at the size of the corrections as we move from NLO 
to NNLO. The up quark distribution at the two orders is illustrated in 
Fig.~\ref{unnlocomps}. As one can see, the change in the central value is 
somewhat larger than the uncertainty due to the experimental errors. The 
predictions for various physical processes have been calculated. The change 
for quark-dominated processes, such as $W$ and $Z$ production, is not very 
large, i.e. $4\%$ or less \cite{MRSTNNLO}, but is sometimes 
bigger than the quoted uncertainty at each 
order. Changes in gluon dominated quantities, such as $F_L(x,Q^2)$, can be 
much larger \cite{MRSTFL}. Similarly there are implications that resummations 
may have significant effects on LHC predictions, particularly at high rapidity
\cite{MRSTerror2}.

\begin{figure}
\vspace{-0.2cm}
\centering\hspace{-0.8cm}\includegraphics[width=.38\linewidth]
{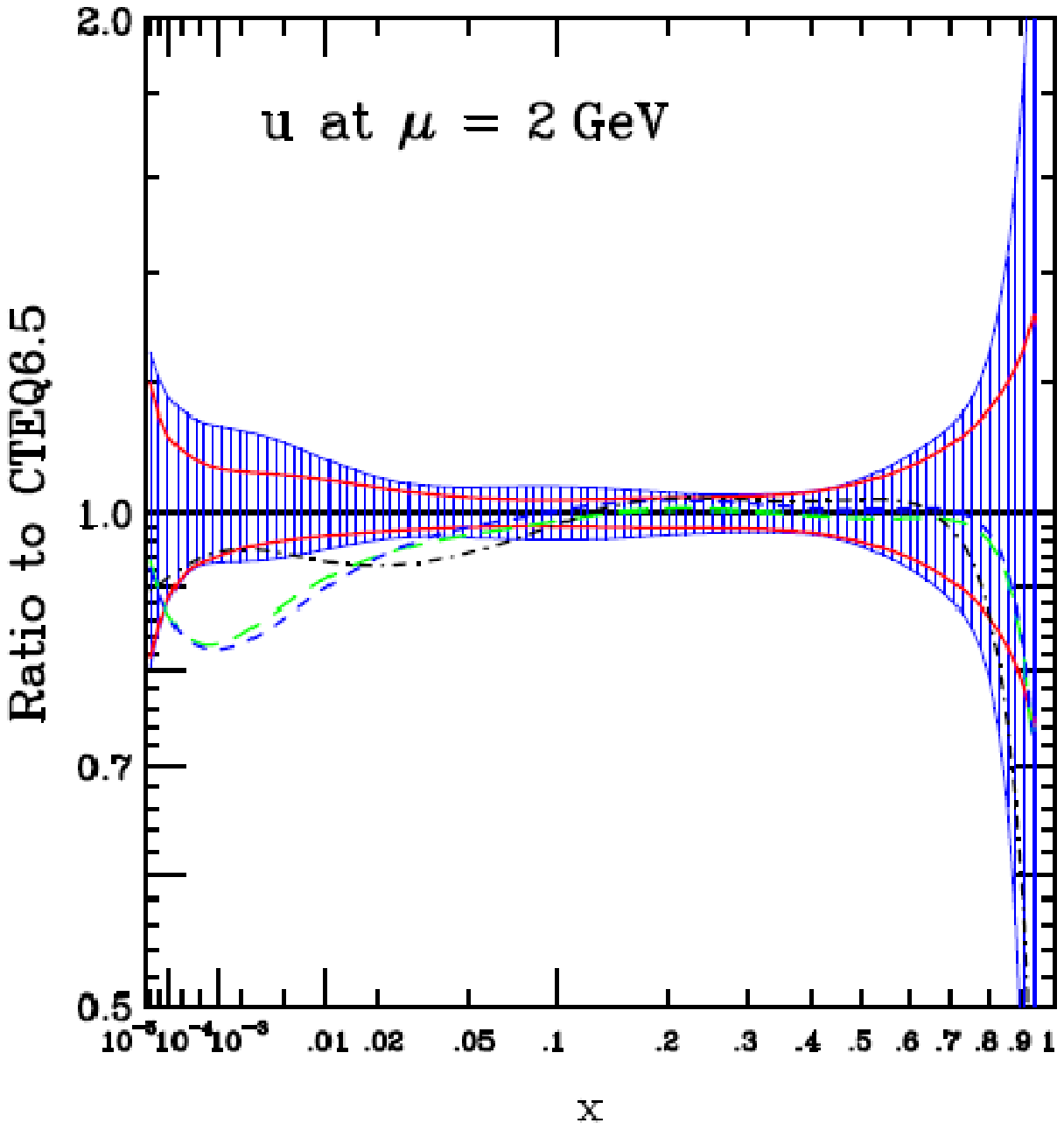}
\hspace{0.2cm}\includegraphics[width=.64\linewidth]{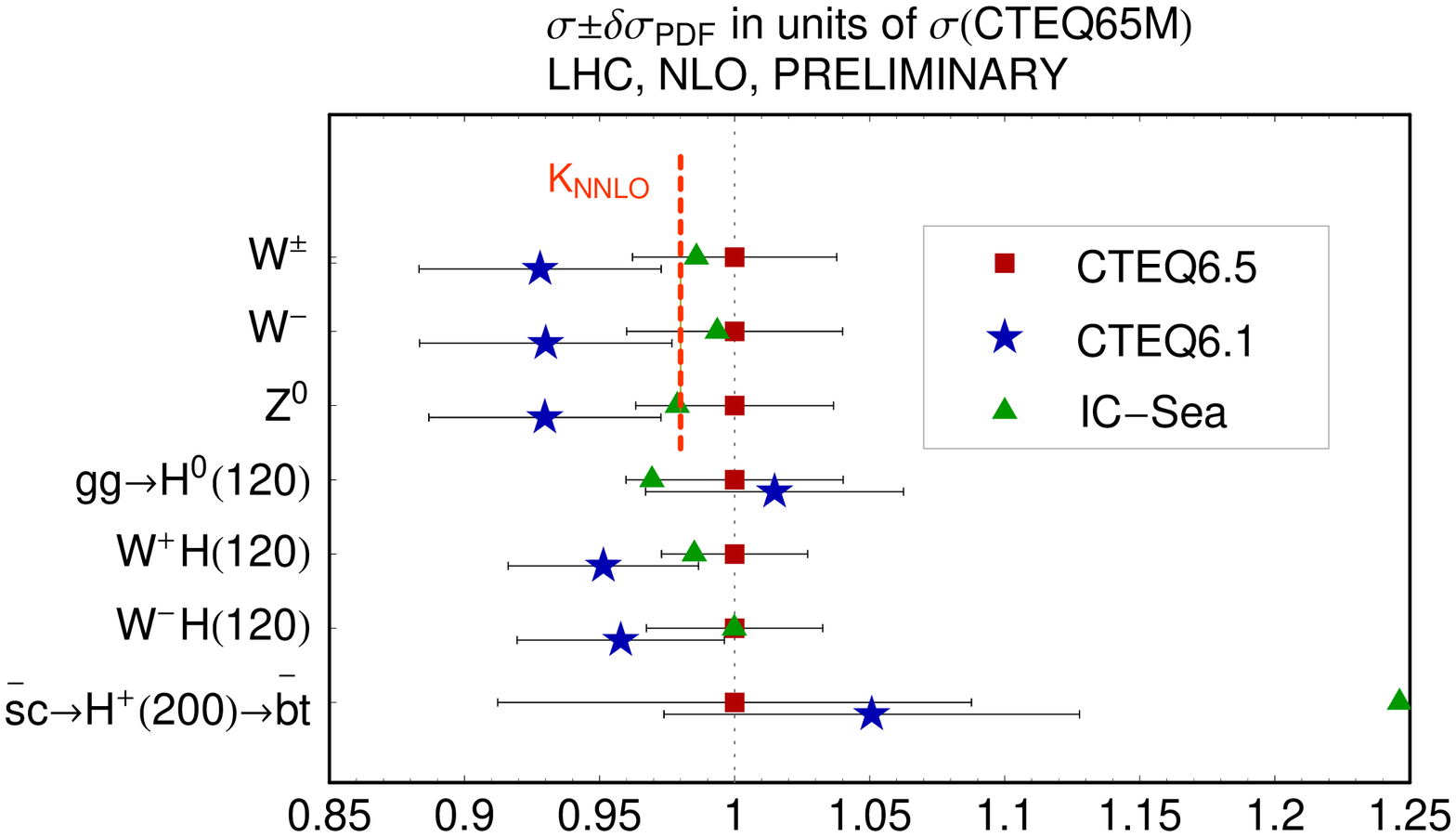}   
\vspace{-0.5cm}
\caption{The CTEQ6.5 up quark with uncertainties compared with previous 
versions, e.g. CTEQ6 (dashed) and MRST04 (dot-dashed) (left) 
and the change in predictions using CTEQ6.5 
partons for LHC cross-sections as opposed to CTEQ6 (right)} 
  \label{cteq6.5}
\vspace{-0.4cm}
\end{figure}

Very recently it has become clear that a less obvious source of theoretical 
errors can have surprisingly large effects, i.e. the precise treatment of 
heavy quark effects. For many 
years CTEQ have had a procedure for extrapolating from 
the limit where quarks are very heavy to the limit where they are effectively
massless, i.e. a 
general-mass variable flavour number scheme (GM-VFNS) \cite{ACOT}. 
Nevertheless, they have chosen the scheme
where the quark masses are zero as soon as the heavy quark evolution begins,
i.e. zero-mass variable flavour number scheme (ZM-VFNS), to be the default 
parton set. In the most recent analysis \cite{CTEQ65} they have 
switched to the GM-VFNS definition as default and noticed that this has a 
very large effect on their small-$x$ light quark 
distributions, mainly determined by 
fitting to HERA data, where mass corrections are 
important, and on LHC predictions. This is shown in Fig.~\ref{cteq6.5}, where 
one sees the prediction for $\sigma_W$ increase by $8\%$.   

\begin{figure} 
\vspace{-0.4cm}
\centering\hspace{-1.3cm}\includegraphics[width=.45\linewidth]
{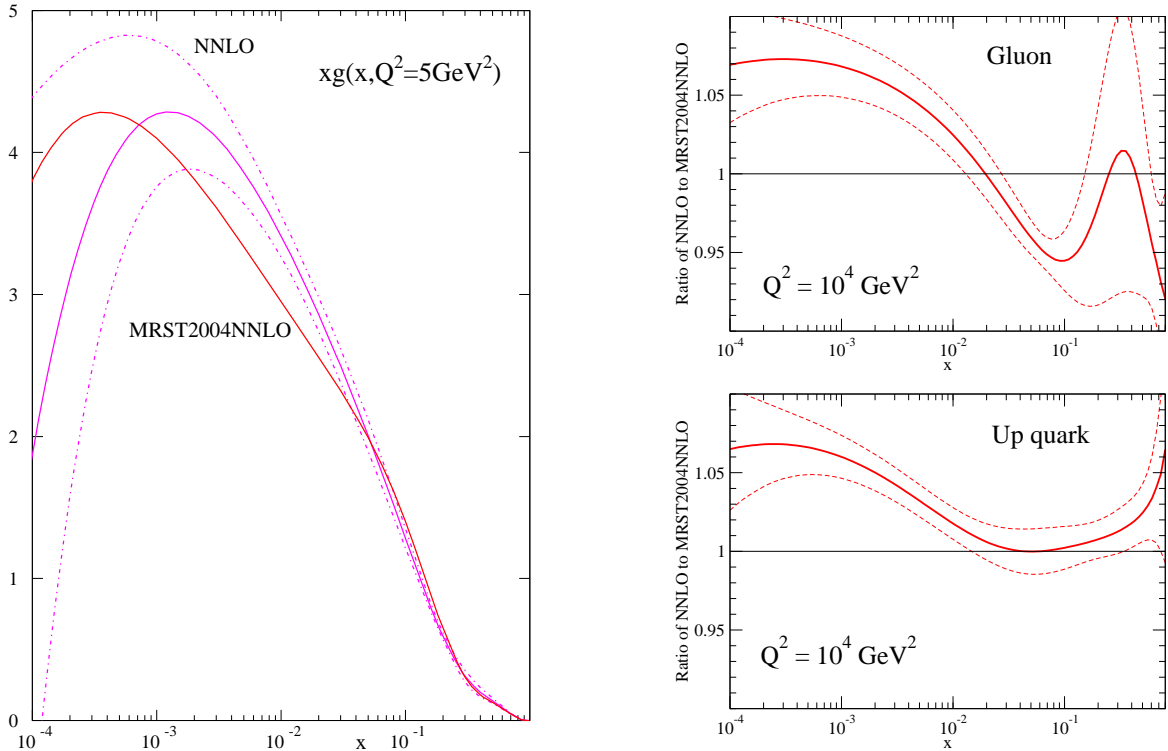}
\hspace{2.3cm}\includegraphics[width=.41\linewidth]{thorne_figure15.eps}   
\vspace{-0.1cm}
\caption{Comparison of the NNLO gluon distribution (and its 
uncertainty) with the previous approximate NNLO distribution at 
$Q^2=5~\GeV^2$ (left), and the ratio at $Q^2=10^4~ \GeV^2$
for both the gluon and the up quark (right).}
  \label{MRST06}
\vspace{-0.1cm}
\end{figure}

Perhaps even more surprising is the change observed by MRST at NNLO. Because 
early approximate ``NNLO'' sets (e.g. \cite{MRSTNNLO}) were based on 
approximate 
splitting functions the MRST group used a (fully explained) approximate 
treatment of heavy quarks at NNLO, in particular not including the 
discontinuities at transition points that occur at this order \cite{buza}.
The correction of this approximate NNLO VFNS between \cite{MRST04} and 
\cite{MRST06} using the scheme in \cite{nnlovfns} led to large 
corrections to the gluon distribution at small $x$ and by evolution, also
to the light quark distributions at higher scales, as seen in 
Fig.~\ref{MRST06}. This results in the corrections to LHC cross-sections shown
in Table 1, i.e. up to $6\%$. In this case the change in procedure was less
dramatic than that for the CTEQ6.5 result, 
where the original approximation was 
of massless quarks, and was also at one order lower. The size of the 
change was certainly unexpected. It is important to note that 
in both these cases the change is not really 
representative of an uncertainty, since each represents a correction of 
something that was known to be wrong. However, in each case the ``wrongness''
was thought to be an approximation requiring only a small correction, an 
expectation that was optimistic. Some parton sets currently available are 
still extracted using similar (or worse) ``approximations'', and even in the 
best case the 
limited order of the calculation means that everything is to some extent an
approximation, with the size of the correction being 
by definition uncertain.       

\begin{table}
\begin{center}
\caption{Total $W$ and $Z$ cross-sections multiplied by leptonic branching
ratios at the Tevatron and the LHC, 
calculated at NNLO using the updated NNLO parton distributions. The 
predictions using the 2004 NNLO sets are shown in brackets.}
\label{tab1}
\begin{tabular}{lll}
\hline\hline
&  $B_{l\nu} \cdot \sigma_W ({\rm nb})$ & $B_{l^+l^-}\cdot\sigma_Z ({\rm nb})$
\\
\hline
% & &  \\
Tevatron &  2.727 (2.693)   &  0.2534 (0.2518)      \\
LHC  & 21.42 (20.15)  & 2.044 (1.918)     \\
\hline\hline
    \end{tabular}

\end{center}
\vspace{-0.6cm}
\end{table}

\section{Conclusions}

One can determine the parton distributions from fits to existing data and 
predict cross-sections at the  LHC.  The fit 
quality using NLO or NNLO QCD is fairly good. 
There are various ways of looking at uncertainties
due to the errors on data. 
For genuinely global fits, using $\Delta \chi^2 =1$ is not 
a sensible option due to incompatibility between data sets and possibly 
between data and theory. Uncertainties due to parton distributions from
experimental errors lead to
rather small, $\sim 1-5 \%$ uncertainties for most LHC quantities, 
and are fairly similar for all approaches.    
However, sometimes the central values using different sets differ by more than 
this. The uncertainties from input assumptions, e.g.  
cuts on data, sets used, parameterisations {\it etc.}, are comparable and 
sometimes  larger than statistical uncertainties. In particular, the detail 
of uncertainties on the flavour 
decomposition of the quarks is still 
developing. 

Uncertainties from higher orders/resummation in QCD are significant, and 
electroweak corrections are also potentially large at very high energies. 
At the LHC measurement at high rapidities, e.g. $W, Z$, would be useful 
in testing our understanding of QCD. Our limited knowledge of the 
theory is often the dominant source of 
uncertainty. There has recently been much progress: more processes known at 
NLO, and some at NNLO; improved heavy flavours treatments; developments in
resummations {\it etc.}. In particular, essentially full NNLO parton
distribution determinations are now possible. 
But further theoretical improvements and complementary measurements are 
necessary for a full 
understanding of the best predictions and their uncertainties.

\end{document}